%% file: zw1742.tex
\documentclass[]{aa}
\usepackage[varg]{txfonts}
\usepackage{times,epsfig} 

\newcommand{\chandra}{{\it Chandra}}
\newcommand{\xmm}{{\it XMM-Newton}}

\newcommand{\ergs}{erg s$^{-1}$}
\newcommand{\ergcms}{erg s$^{-1}$ cm$^{-2}$}
\newcommand{\obj}{Zw~1742$+$3306}

\begin{document}

\title{Cold fronts and metal anisotropies \\ in the X-ray cool core of the galaxy cluster \obj}

\author{
S. Ettori \inst{1,2}, F. Gastaldello \inst{3, 4}, M. Gitti \inst{5}, E. O'Sullivan \inst{6}, M. Gaspari \inst{7}, F. Brighenti \inst{5}, L. David \inst{6}, A.C. Edge\inst{8}
}

\institute{ 
INAF, Osservatorio Astronomico di Bologna, via Ranzani 1, I-40127 Bologna, Italy \\ \email{stefano.ettori@oabo.inaf.it}
\and
INFN, Sezione di Bologna, viale Berti Pichat 6/2, I-40127 Bologna, Italy
\and
INAF, IASF,  Via E. Bassini 15, I-20133 Milano, Italy
\and
University of California Irvine, 4129, Frederick Reines Hall, Irvine, CA, 92697-4575, USA
\and
Physics and Astronomy Department, University of Bologna, via Ranzani 1, 40127 Bologna, Italy
\and
Harvard-Smithsonian centre for Astrophysics, 60 Garden Street, Cambridge, MA 02138, USA
\and
Max Planck Institute for Astrophysics, Karl-Schwarzschild-Strasse 1, 85741 Garching, Germany
\and
Department of Physics, University of Durham,  south Road,  Durham, DH1 3LE, UK
}

\authorrunning{Ettori et al.} 
\titlerunning{The X-ray cool core in \obj}

\date{Accepted on 14 May 2013}

\abstract
{In recent years, our understanding of the cool cores of galaxy clusters has changed.
Once thought to be relatively simple places where gas cools and flows toward the centre, now they are believed to be very dynamic places where heating from the central Active Galactic Nucleus (AGN) and cooling, as inferred from active star formation, molecular gas, and H$\alpha$ nebulosity, find an uneasy energetic balance.} 
{We want to characterize the X-ray properties of the nearby cool-core cluster \obj, selected because it is bright at X-ray (with a flux greater than $10^{-11}$ \ergcms\  in the 0.1--2.4 keV band) and $\rm{H}\alpha$ wavelengths ($\rm{H}\alpha$ luminosity $> 10^{40}$ \ergs).}
{We used \chandra\ data to analyse the spatial and spectral properties of the cool core of \obj, a galaxy cluster at $z=0.0757$ that emits in $H\alpha$ and presents the brightest central galaxy located in a diffuse X-ray emission with multiple peaks in surface brightness.}
{We show that the X-ray cool core of the galaxy cluster \obj\ is thermodynamically very active with evidence of cold fronts and a weak shock in the surface brightness map and of an apparently coherent, elongated structure with metallicity greater than the value measured in the surrounding ambient gas by about 50\%. This anisotropic structure is $280 \times 90$ kpc$^2$ and is aligned with the cold fronts and with the X-ray emission on larger scales. We suggest that all these peculiarities in the X-ray emission of \obj\ are either a very fine-tuned output of a sloshing gas in the cluster core or the product of a metal-rich outflow from the central AGN.}
{}

\maketitle

\section{Introduction}

\begin{figure*}
\epsfig{figure=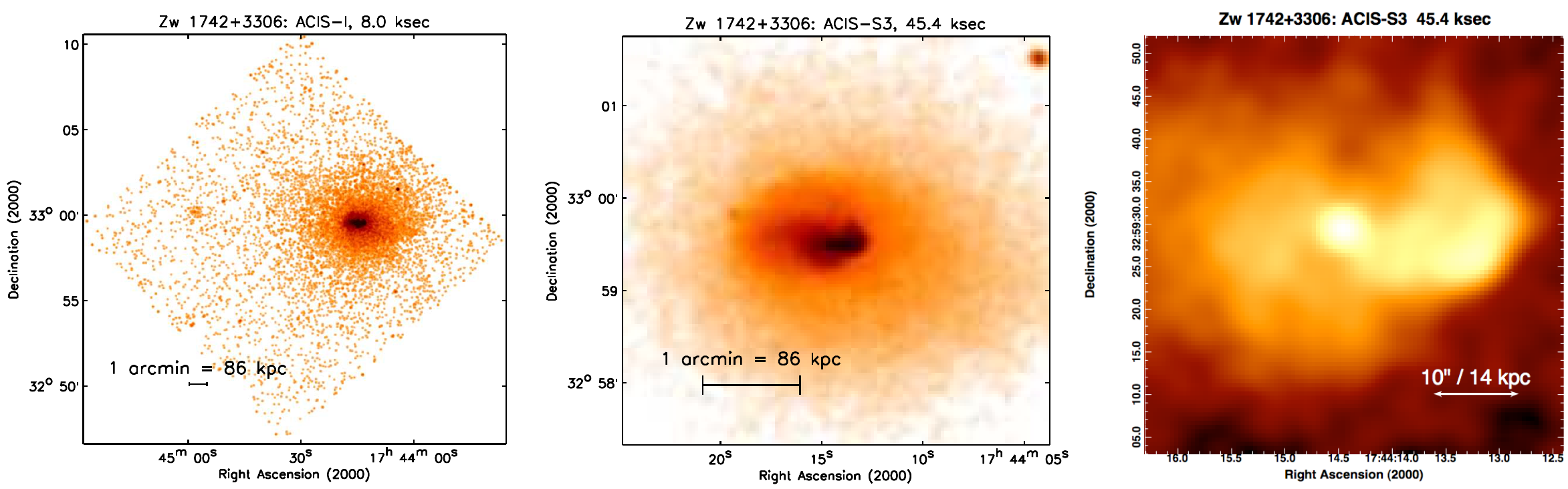,width=\textwidth}
\caption{ACIS-I ({\it left}) and ACIS-S3 ({\it centre}) exposures of \obj. The images have a bin-size of 1.968 arcsec and have been convolved with a Gaussian with FWHM of 3 pixels.
{\it Right}: a zoomed version of the ACIS-S3 image with bin-size of 0.492 arcsec and convolved with a Gaussian of 3 sigma in pixel.
} \label{fig:image} \end{figure*}

A considerable fraction of the local population of clusters of galaxies
($\sim$ 35\%, Eckert et al. 2011 and references therein) is characterized by a prominent
surface brightness peak, usually coincident with
the centre of the large scale X-ray isophotes and with the position
of the brightest central galaxy (BCG), associated with a
decrease of the temperature profile in the inner regions and a
positive gradient of the metal abundance profile. 
In the central regions of these cool core clusters, the gas has a radiative cooling time short enough
that it should cool, condense, and flow toward the centre with mass cooling rates 
as high as $10^2$-$10^3 M_{\sun}$  yr$^{-1}$. In the absence of a heating source, the gas 
will accrete onto the BCG and form stars (Fabian 1994). 
The end products of cooling in the forms of cold molecular clouds and star formation are observed in 
many BCGs (e.g. Edge 2001, O'Dea et al. 2008) but at a level of at least an order of magnitude below
those expected from uninterrupted cooling over the age of clusters.
X-ray observations with \chandra\ and \xmm\ provided crucial evidence against a simple cooling flow model
establishing a lack of strong emission from gas cooling below $\sim T_{\rm vir} / 3$, indicating 
much less cool gas than expected in a steady cooling flow and thus lowering the previous cooling rates
by a factor of 5-10 (e.g. Peterson \& Fabian 2006).
A compensating heat source must therefore resupply the radiative losses and quench the cooling flow.
The most likely heat source is mechanical heating from the radio Active Galactic Nucleus (AGN) in the BCG at the centre of the cool core.
The new established paradigm therefore involves a repetitive feedback cycle where the massive black hole in the central
galaxy, fed by the gas also forming molecular clouds and stars, launches mechanically powerful radio jets / outflows
that heat the surrounding atmosphere, slowing efficiently the rate of cooling.
There is clear observational evidence for the steps involved in the feedback process, since 
almost all cool cores harbour radio sources (Burns 1990, Sun 2009) that produce bubbles of relativistic plasma 
inflated by the radio jets. These jets interact with the surrounding ICM creating X-ray surface brightness depressions (or cavities) 
causing heating as they rise through the ICM. Together with weak shocks associated with the outburst, the energies provided by the AGN are 
indeed comparable to those needed to prevent gas from cooling. Many excellent reviews of this topic are now available 
(Fabian 2012, McNamara \& Nulsen 2012, Gitti et al. 2012).

A distinctive feature of cool cores is the presence of an iron abundance peak, higher on average by a factor of 3 to 4 than the bulk 
of the ICM, consistent with being produced by stars associated with the BCG (e.g. De Grandi \& Molendi 2001, De Grandi et al. 2004).
Several studies have shown that the iron abundance profile is broader than the stellar light profile of the BCG 
(e.g. Rebusco et al. 2006, Graham et al. 2006) indicating that metals are diffusing outward, their distribution being broadened
by mergers or AGN outflows, even though allowance should be made for the opposite process of concentration of optical light naturally 
associated to the formation of the BCG (De Grandi \& Molendi 2001). 
AGN bipolar outflows have recently received strong observational and theoretical support: 
following early results showing metal-rich gas along the cavities and radio jets of some individual 
clusters and groups (e.g. Simionescu et al. 2008, 2009; Gitti et al. 2011; Kirkpatrick et al. 2009; O'Sullivan et al. 2011),
Kirkpatrick et al. (2011) found consistently in a sample of ten clusters an anisotropic distribution of iron aligned with the cavity 
and large scale radio emission axes. The radial extent of the metal outflow is found to scale with the jet power and 
it is greater than the extent of the inner cavities, showing this to be a long-lasting effect sustained over multiple generations of outbursts. 
The amount of transported gas is substantial, and it is consistent with the results of simulations showing that AGN outflows are able to advect ambient, 
iron-rich  material from the core to much larger radii of the order of hundreds of kpc (e.g. Gaspari et al. 2011a, 2011b).  

Another striking feature revealed by \chandra\ and \xmm\ in cool cores is the almost ubiquitous presence of cold fronts 
(e.g. Ghizzardi et al. 2010). Cold fronts are sharp surface brightness discontinuities, interpreted as contact edges between 
regions of gas with different entropies (see the review by Markevitch \& Vikhlinin 2007).
In cool core clusters, cold fronts are most likely induced by minor
mergers that produce a disturbance on the gas in the core of the main cluster, displace it from the
centre of the potential well, and decouple it from the underlying dark matter halo through ram
pressure (e.g. Ascasibar \& Markevitch 2006).
The oscillation of the gas of the core around the minimum of the potential generates
a succession of radially propagating cold fronts, appearing as concentric edges in the surface
brightness distribution of the cluster. These fronts may form a spiral structure when the sloshing
direction is near the plane of the sky and the merger has a non-zero angular momentum. When the
sloshing direction is not in the plane of the sky, concentric arcs are observed.
The sequence of events is described in great detail in the simulations presented in
Ascasibar \& Markevitch (2006) and Roediger et al. (2011).

Last but not least, observational signature of cool cores is the fact that the BCGs at their centres
have extensive optical emission-line (H$\alpha$) luminosity (e.g. Crawford et al. 1999, McDonald et al. 2010). The emission can extend
for tens of kpc and in the most extreme case of NGC~1275 in Perseus out to 100 kpc. The ionization mechanism
of these filaments is still much debated; however, the observational evidence increasingly points to their
origin in the cooling ICM, as the strength of the H$\alpha$ emission correlates with the intensity of blue continuum
due to star formation and to the amount of molecular gas (e.g. Crawford et al. 1999, Edge 2001).
The relation between short cooling times, star formation, and H$\alpha$ emission has been resolved in a sharp threshold:
high star formation rates and nebular emission are preferentially found in BCGs in atmospheres with central cooling times
shorter than 0.5 Gyr (or equivalently with central entropies lower than 30 keV cm$^2$, Rafferty et al. 2008; Cavagnolo et al. 2008).
These phenomena can all be linked by the occurrence of thermal instabilities, with warm gas condensing out of the hot keV phase
and forming layers of H$\alpha$ surrounding the dense core of molecular gas (e.g. Gaspari et al. 2012 and references therein).
Clearly H$\alpha$ emission is a tracer of cooling activity and it has been used to indirectly probe the evolution of cool cores with time 
(McDonald 2011, Samuele et al. 2011). 
This connection can be also exploited locally to select additional interesting objects to study the cooling/heating balance in cool cores.
This is the case of the galaxy cluster analysed in the present work, \obj, a nearby cool-core cluster with H$\alpha$ emission associated with it.

\begin{figure*}
\hbox{
  \epsfig{figure=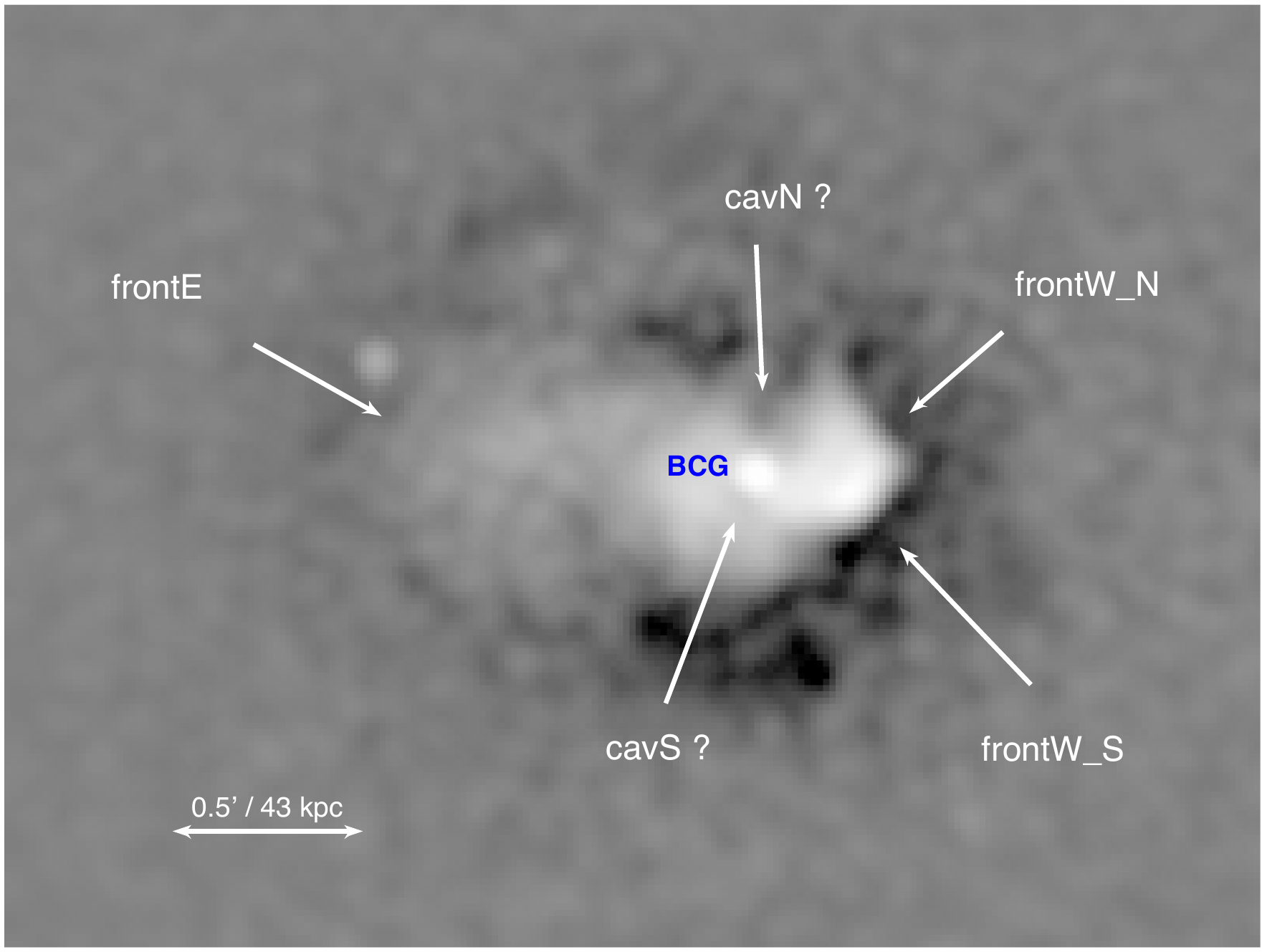,width=0.5\textwidth}
  \epsfig{figure=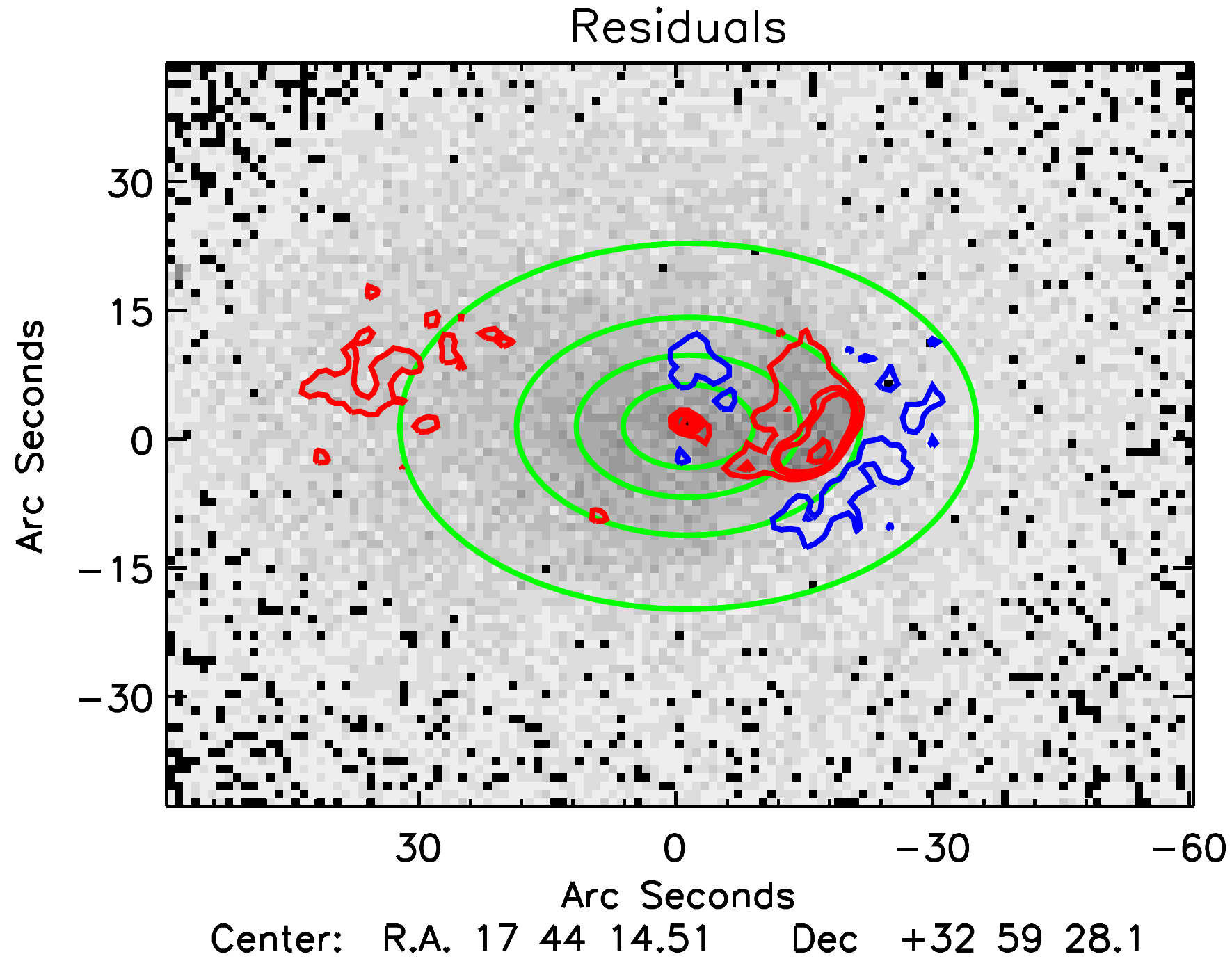,width=0.5\textwidth}
}
\caption{Deviations from the smoothed X-ray emission.
({\it Left}) Unsharp mask obtained by subtracting two Gaussian convolved images (with FWHM of 2 and 20 pixels). 
Some of the features discussed in the text are shown.
({\it Right}) Exposure-corrected X-ray image overplotted with the best-fit 2D $\beta-$model (green contours) and the positive (red) and negative (blue) residuals at 
68.3\%, 90\%, 95\%, and 99.73\% level of confidence. 
} \label{fig:unsharp} 
\end{figure*}

\begin{figure*}
\centering
\epsfig{figure=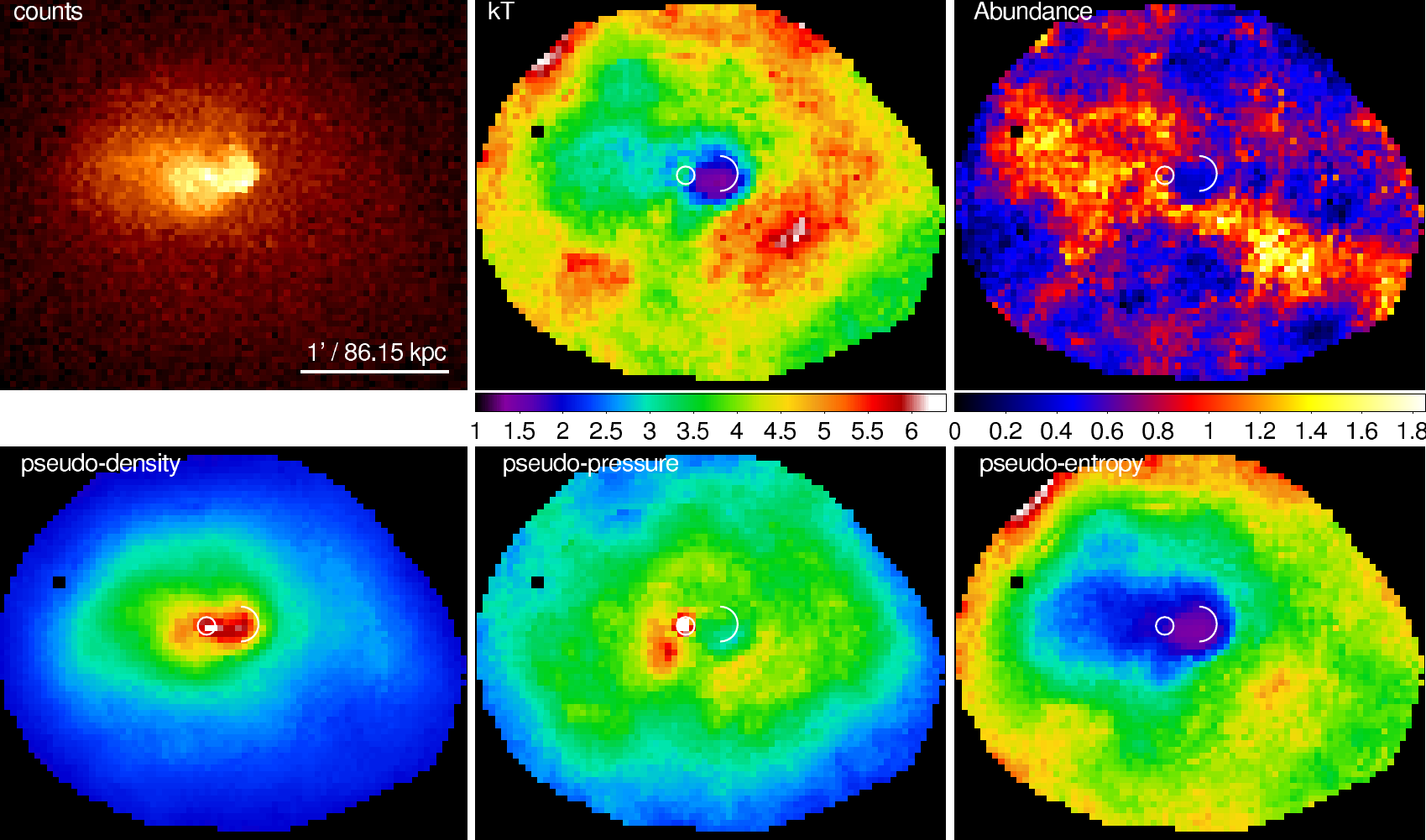,width=0.9\textwidth}
\caption{0.5-7 keV counts image of \obj\ compared with maps of temperature,
abundance, pseudo-density, pseudo-pressure, and pseudo-entropy. Map pixels
are about $2.5''$  in size, and correspond to circular spectral extraction regions
of radius $4-20''$ in radius containing $>$ 1500 net counts. Uncertainties on
temperature, density, and pressure/entropy are less than 10 \%, 5 \%, and 11 \%, respectively.
Abundance ranges from 0.15 (purple) to $\sim$1.5 (red) solar, but
uncertainties on abundance are large, generally $\sim$Ä50\%. 
The apparent low abundances associated with the coolest region are probably caused by
multi-temperature gas along the line of sight.
The BCG (white circle) and the front to the west (white arc) are indicated.
} \label{fig:specmap} \end{figure*}

\section{Properties of \obj} 

We have selected \obj, also known as Z8276 or PLCKESZ~G057.92+27.64, 
as a source detected with $S/N = 6.12$ in the {\it Planck} Early Sunyaev-Zeldovich catalogue (Planck Collaboration 2011), 
from the ROSAT Brightest Cluster Sample (BCS) sample (203 clusters in the northern hemisphere, 90\% complete at $4.4 \times
10^{-12}$ \ergcms, Ebeling et al. 1998) among the objects with X-ray
fluxes greater than $10^{-11}$ \ergcms\ in the 0.1--2.4 keV band and
H$\alpha$-lumininosity $> 10^{40}$ \ergs\ as quoted in
Crawford et al. (1999). Most of the objects satisfying this selection
are well-known cool core clusters (i.e. ranked according to their
X-ray flux: A2199, A1795, A2052, A478, A2634, A2204; \obj\ is the seventh in this list)
and have been early tests of the AGN paradigm.

The galaxy cluster \obj\ is located at redshift 0.0757. One arcmin (arcsec) corresponds, therefore, to 86.15 (1.44) kpc for the
cosmological parameters adopted throughout this work 
($H_0=70$ km s$^{-1}$ Mpc$^{-1}$, $\Omega_{\rm m}=1-\Omega_{\Lambda}=0.3$). 
It appears in the BCS with an X-ray luminosity
of 2.34$\times 10^{44}$ erg~s$^{-1}$ (Ebeling et al. 1998).
The brightest cluster galaxy at (RA, Dec) = (17$^h$ 44$^m$ 14.5$^s$,
+32$^\circ$ 59$^m$ 30$^s$) has an H$\alpha$ luminosity of
9.7$\times 10^{40}$ erg~s$^{-1}$ and analysis of the optical
continuum implies a star formation rate of 0.3 M$_\odot$~yr$^{-1}$
(Crawford et al. 1999). The BCG has a total infrared luminosity
of 7.4$\times 10^{43}$ erg~s$^{-1}$ derived from {\it Spitzer}
IRAC amd MIPS photometry (Quillen et al. 2008) that corresponds
to an equivalent star formation rate of 3.7 M$_\odot$~yr$^{-1}$
(O'Dea et al. 2008). The discrepancy between the optical
and MIR star formation rates can be accounted for by dust
absorption in the optical. Finally, the BCG contains a bright (74mJy
at 5GHz, Gregory \& Condon 1991), flat spectrum ($\alpha\sim-0.5$)
radio source, but is also associated with a steeper spectrum
component ($\alpha\sim-1.2$) detected at 74 and 325MHz in 
VLSS (Cohen et al. 2007) and WENSS (Rengelink et al. 1997). 
Without higher resolution radio imaging, the relationship between these
two components remains unclear.

We present here the results obtained from our recent \chandra\ ACIS-S observation (PI: Gitti; nominal exposure of 44 ksec performed in November 2009).
We also discuss the 8 ksec ACIS-I exposure (PI: Murray; collected in January 2007) available in the archive.

\section{X-ray spatial and spectral analysis}

The ACIS-S exposure was reprocessed with the most recent calibration files at the moment of the data analysis ({\tt CALDB 4.4.10}). We applied the prescriptions recommended for ACIS data reduction using the {\tt CIAO 4.4} package. No flares are evident in the light-curve and the good-time intervals cumulate to a total of 45.4 sec. The ACIS-I exposure was also reprocessed allowing us to use 8.0 ksec out of the 8.4 ksec archived.
The latter dataset was used only to produce images.
Hereafter, a Galactic absorption corresponding to a column density $n_H$ of $3.83 \times 10^{20}$ particle cm$^{-2}$ is assumed.

In Fig.~\ref{fig:image}, we show the raw image obtained in the soft (0.5--2 keV) band. The unsharp mask image, obtained by subtracting the raw image 
convolved with a Guassian of $2''$ and $20''$, is shown in Fig.~\ref{fig:unsharp} with some interesting features overplotted.
We also fitted a 2D $\beta-$model, with a fixed centre on the cD galaxy. We obtained the best-fit parameters
$r_{\rm c} = 15.9 \pm 0.5$ arcsec, $\beta = 0.48 \pm 0.01$, ellipticity $\epsilon = 0.37 \pm 0.01$, and angle of ellipticity of $0\pm1$ degree.
The residuals after the subtraction of the 2D $\beta-$model from the raw image are shown in Fig.~\ref{fig:unsharp}.
The central AGN is clearly detected as a point source in X-rays.
This source seems rather stable in the two Chandra observations
(2007-01-28 for the ACIS-I observation,  2009-11-26 for the ACIS-S observation).
A fit with a power law (and fixed Galactic absorption) provides
a constraint on the photon index of $1.7 \pm 0.1$ and estimates an un-absorbed flux of $4.8 \times 10^{-14}$ \ergcms\ 
and $9.3 \times 10^{-14}$ in the 0.5--2 and 2--10 keV bands, respectively, corresponding to an X-ray luminosity of
$6.6 \times 10^{41}$ \ergs\ and  $1.3 \times 10^{42}$ \ergs.

The X-ray centroid, estimated after the exclusion of the all the detected point-sources (BCG included) over a region with an increasing radius up to 200 arcsec from the BCG, oscillates around the brightest central galaxy, moving to the east by 3 arcsec and, then, to the west by about 4 arcsec.
Over larger scales, we modelled the extended X-ray emission enclosing 60\%, 70\%, 80\%, and 90\% of the X-ray light with an ellipse with ellipticity (equal to $1-b/a$, where $b$ and $a$ are the minor and major axis, respectively) varying from 0.29 to 0.14 and an angle from 175 to 165 degrees (measured anti-clockwise from a Cartesian X-axis), confirming the elliptical, irregular shape of the X-ray surface brightness.

For the spectral analysis, the cluster's X-ray emission was modelled with both one and two thermal models ({\tt apec} model in {\tt XSPEC 12.7.1}). The spectral fit was performed in $0.6-7$ keV after the subtraction of a background obtained from the blank-fields exposures renormalized by the particle background emission in the cluster's exposure in the the 9.5--12 keV band and allowing for some soft contribution estimated by comparing spectra in the 0.4--1 keV band extracted from the blank-fields and from the cluster's field free from ICM emission. The results of this spectral analysis for the regions described below are shown in Table~\ref{tab:spectra}. 

In Fig.~\ref{fig:specmap}, we present the results from spectral mapping of the cluster
core. Maps were created using the techniques described in O'Sullivan et
al. (2011), with each pixel representing the best-fitting
values from absorbed {\tt apec} model fits to spectra with 1500 net counts.
Since the spectral extraction regions are typically larger than the map pixels, individual pixel values are not
independent and the maps are analogous to adaptively smoothed images, with
more smoothing in regions of lower surface brightness. To investigate the
2D variation in entropy and pressure in the ICM, we follow the method of
Rossetti et al. (2007) and estimate a pseudo-density,
based on the square root of the normalization of the {\tt apec} model divided by
the area of the spectral extraction region. This is proportional to the
integrated electron density along the line of sight. The maps of
pseudo-density, pseudo-entropy and pseudo-pressure should thus trace any projected
structure in these important quantities. 

\begin{figure*}
\hbox{
 \epsfig{figure=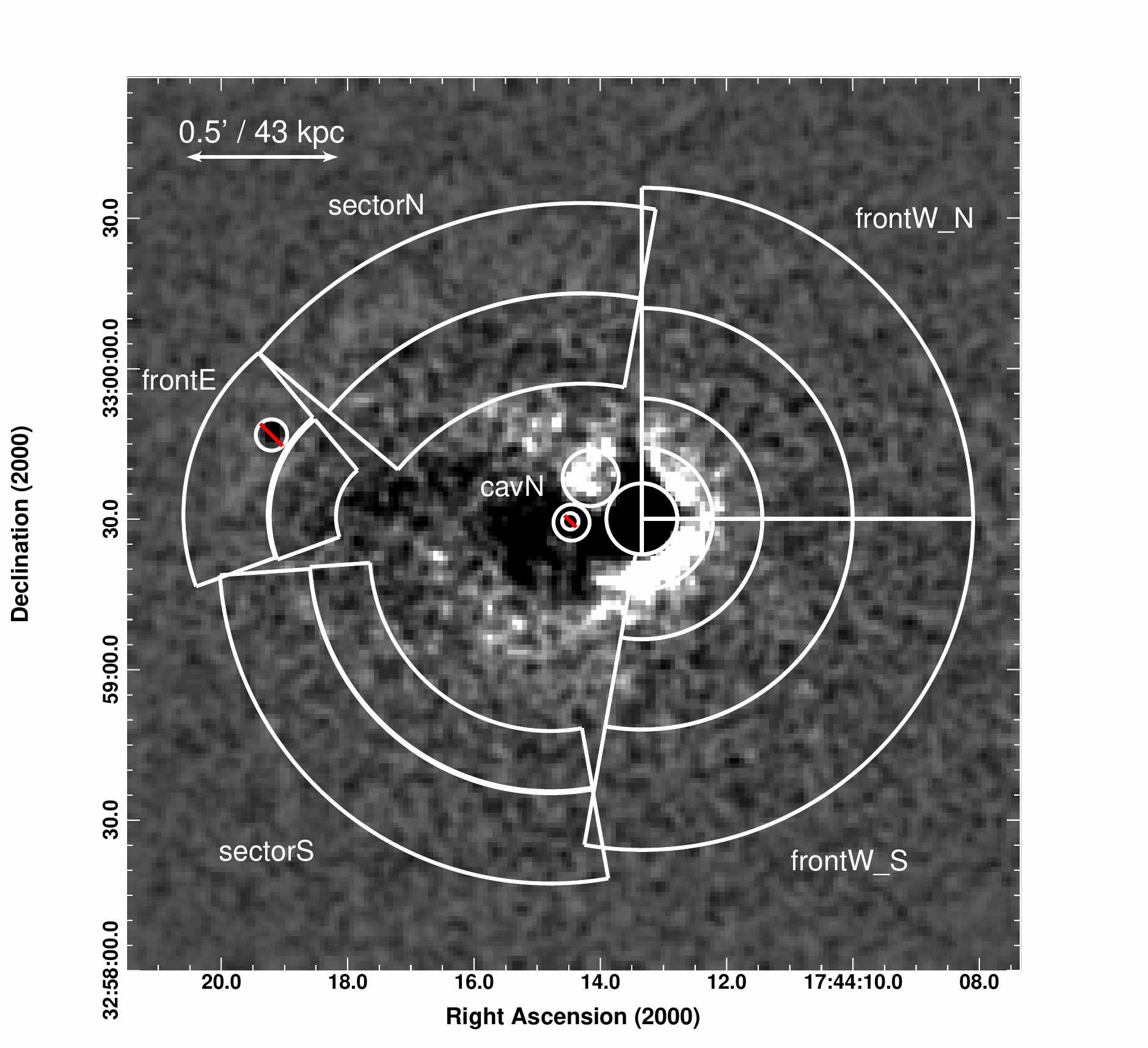,width=0.5\textwidth}
 \epsfig{figure=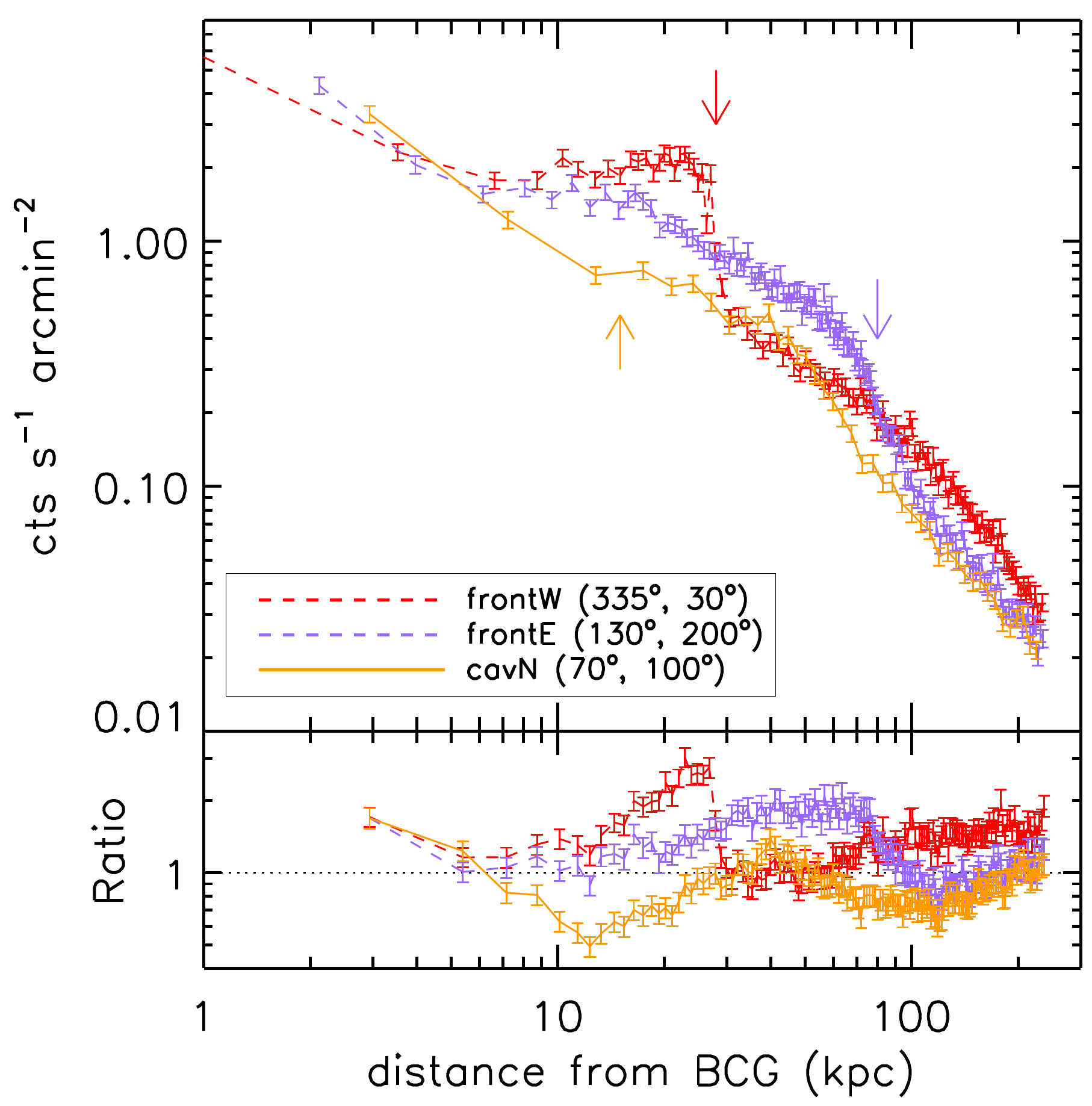,width=0.5\textwidth}
} \caption{({\it Left}) Regions of interest for the spectral analysis (see Table~\ref{tab:spectra}) overplotted to  the unsharp mask: moving outwards,
{\it sectorN} and {\it sectorS} are divided in two regions ({\it in} and {\it out}); {\it frontE} is divided in 2 regions across the front;
{\it frontW-in} indicates the circular region in front of the edge; {\it frontW-inN} and {\it frontW-outN, ..., out4N} are the regions across the front 
pointing to the north-west; {\it frontW-inS} and {\it frontW-outS, ..., out4S} are the regions pointing to south-west.
({\it Right}) Surface brightness profiles extracted from the exposure-corrected 0.5--2 keV image fixing the centre on the BCG. 
The angles that define the sectors are measured anti-clockwise from the positive X-axis (angle $=0^{\circ}$).
The arrows indicate the positions of the two fronts where there is a sharp steepening of the gradient.
At the bottom, the ratios between the surface brightness profiles extracted in sectors and the profile obtained from the remaining X-ray emission assumed to be more undisturbed are plotted. 
} \label{fig:mapt} \end{figure*}

\begin{figure*}
\hbox{
\epsfig{figure=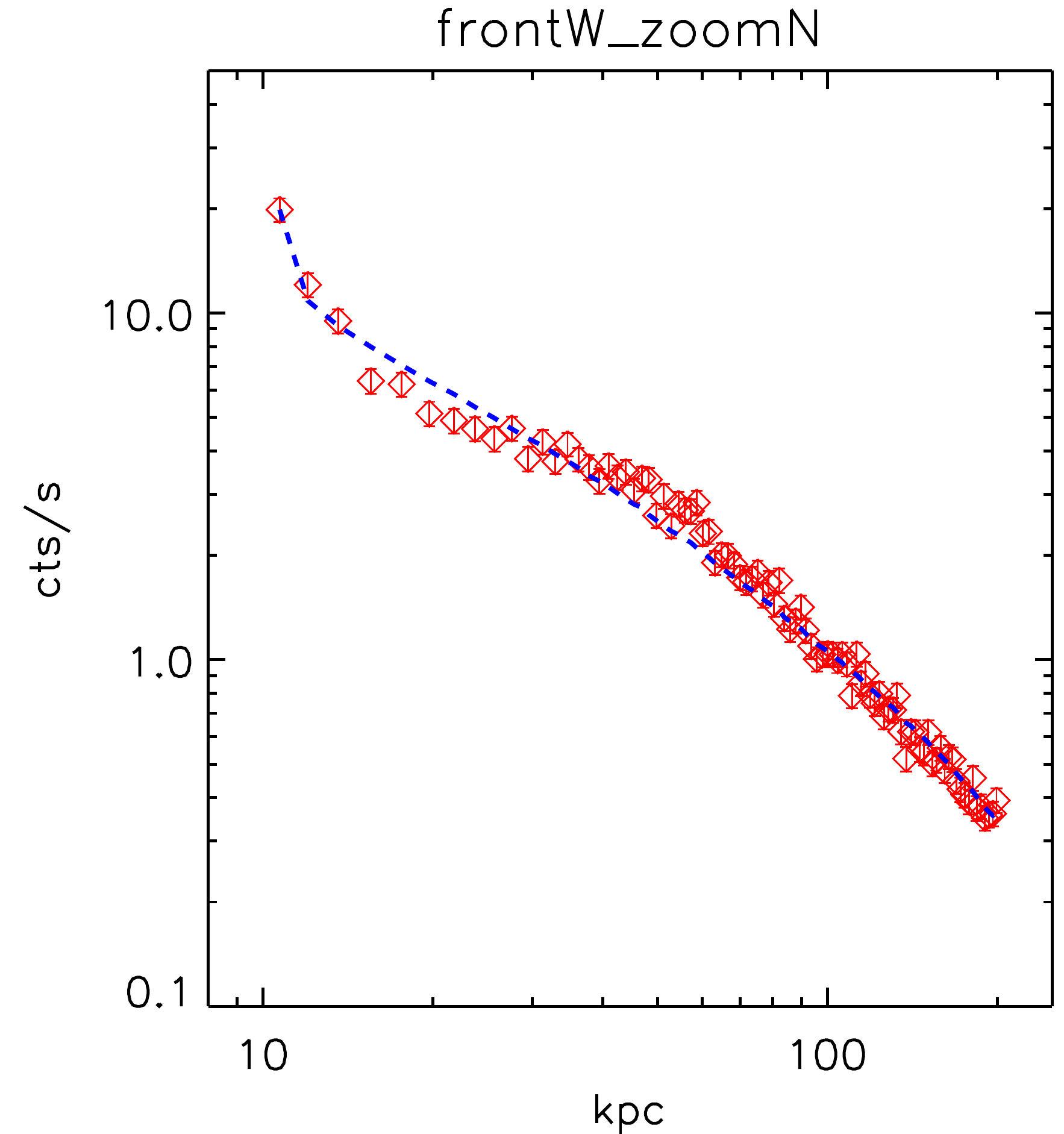,width=0.25\textwidth}
\epsfig{figure=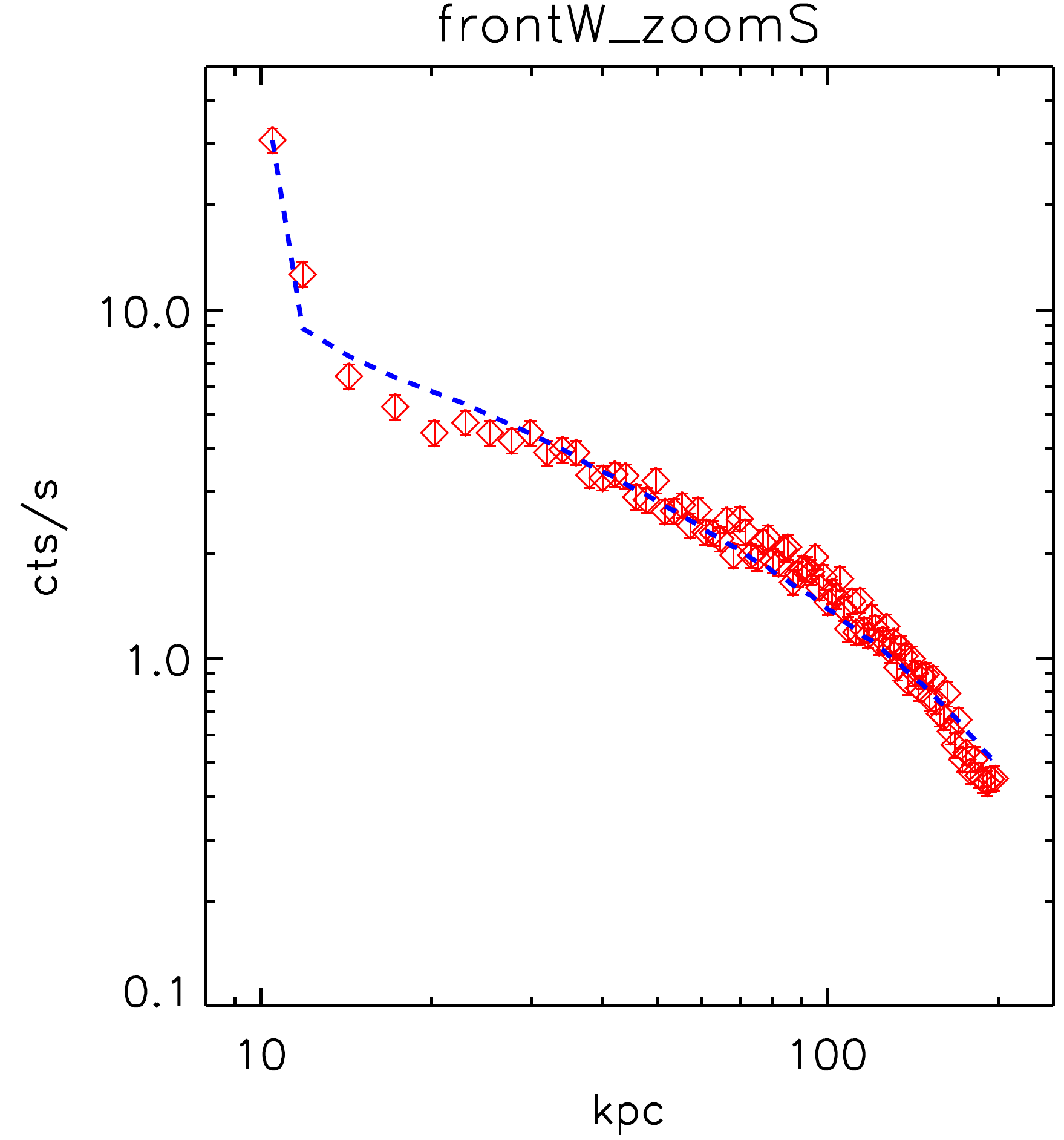,width=0.25\textwidth}
\epsfig{figure=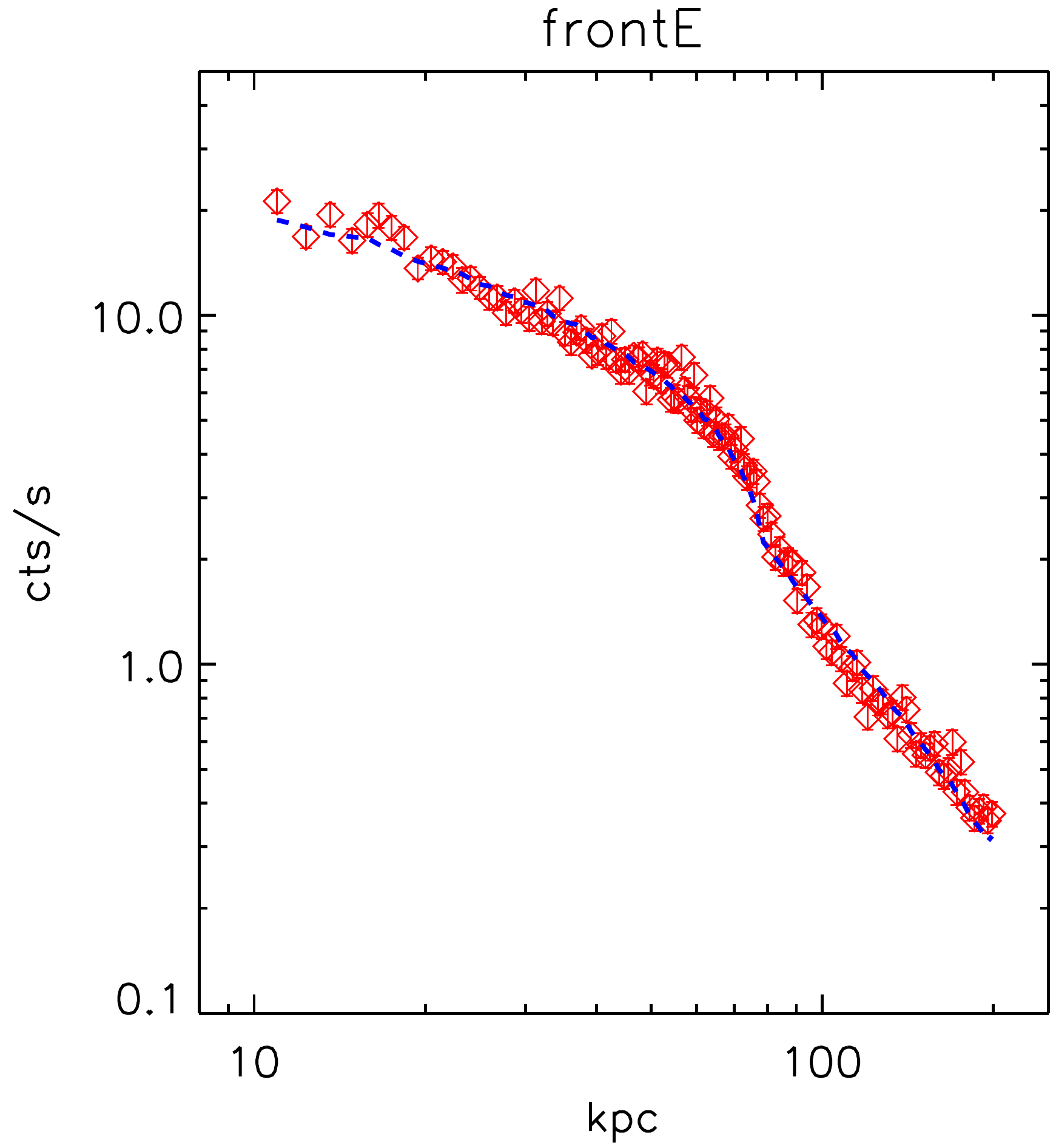,width=0.25\textwidth}
\epsfig{figure=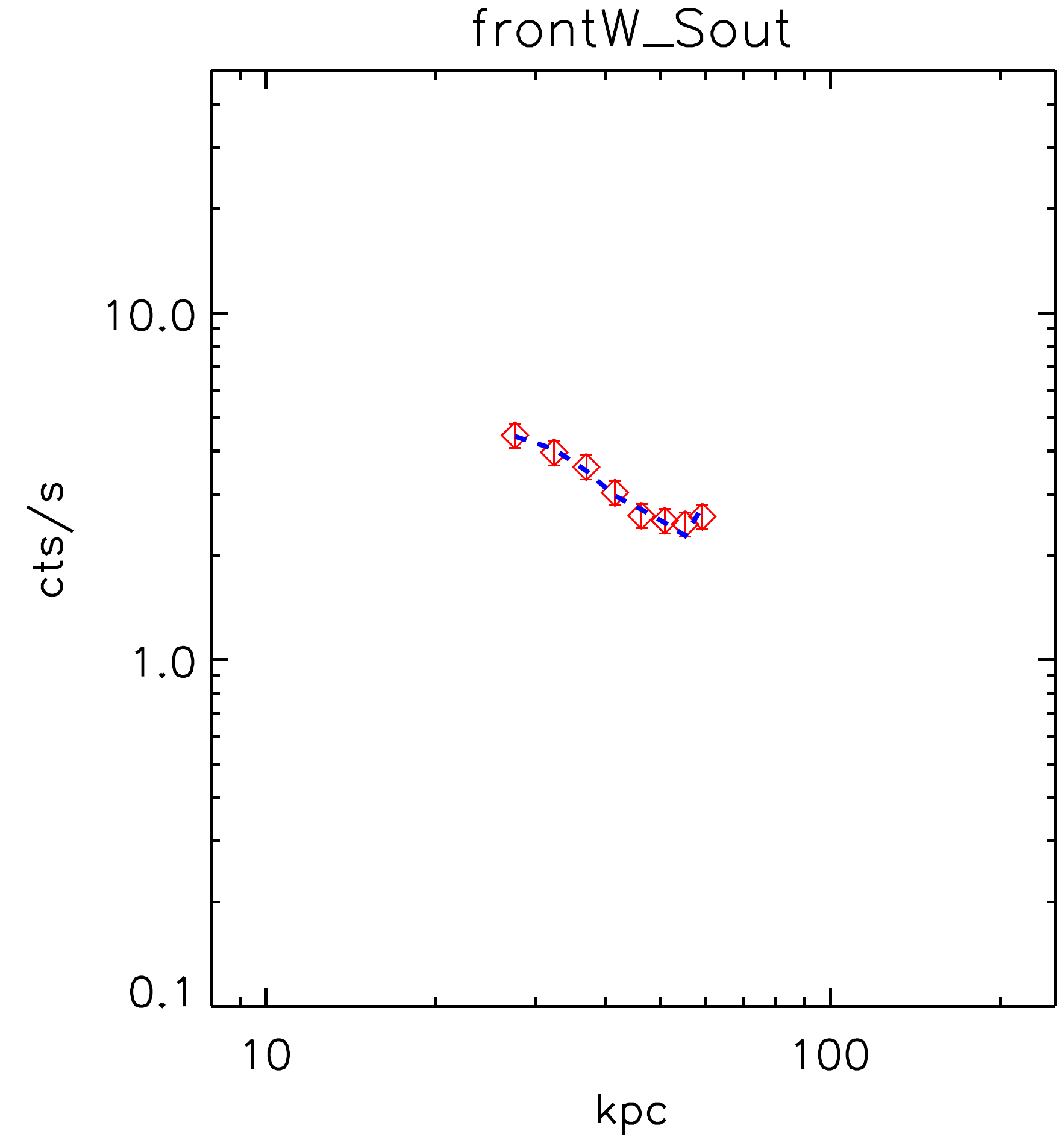,width=0.25\textwidth}
} \caption{Surface brightness profiles of the fronts detected in our analysis with the best-fit with a broken-power law describing the gas density profile
overplotted with a dashed line.
 ({\it From left to right}): to the west, northern and southern sectors, at about 12 kpc from the centre fixed at (RA, Dec) = (17$^h$ 44$^m$ 13.0$^s$, +32$^\circ$ 59$^m$ 22$^s$) and 30 kpc from the BCG; to the east, at 80 kpc from the centre located on the BCG; to the south/west at about 60 kpc from the BCG. 
The centre of the annular regions used to extract each profile has been chosen to best match the curvature of the radial bins with the shape of the edges as seen in Fig.~\ref{fig:mapt}.
 } \label{fig:fronts} \end{figure*}

Combined with the unsharp mask,
these maps allow us to identify a number of interesting features, which
are further characterized through surface brightness profiles extracted in
circular annuli from the exposure-corrected 0.5-2 keV image, centreed on
the BCG (see Fig.~\ref{fig:mapt}). 
As we discuss in the following sections, we find robust indication 
for a cold front at $\sim 16''$ (23 kpc) west of the BCG
and a front at about 80 kpc to the east  (see Fig.~\ref{fig:fronts}). 

\begin{figure}
\centering
 \epsfig{figure=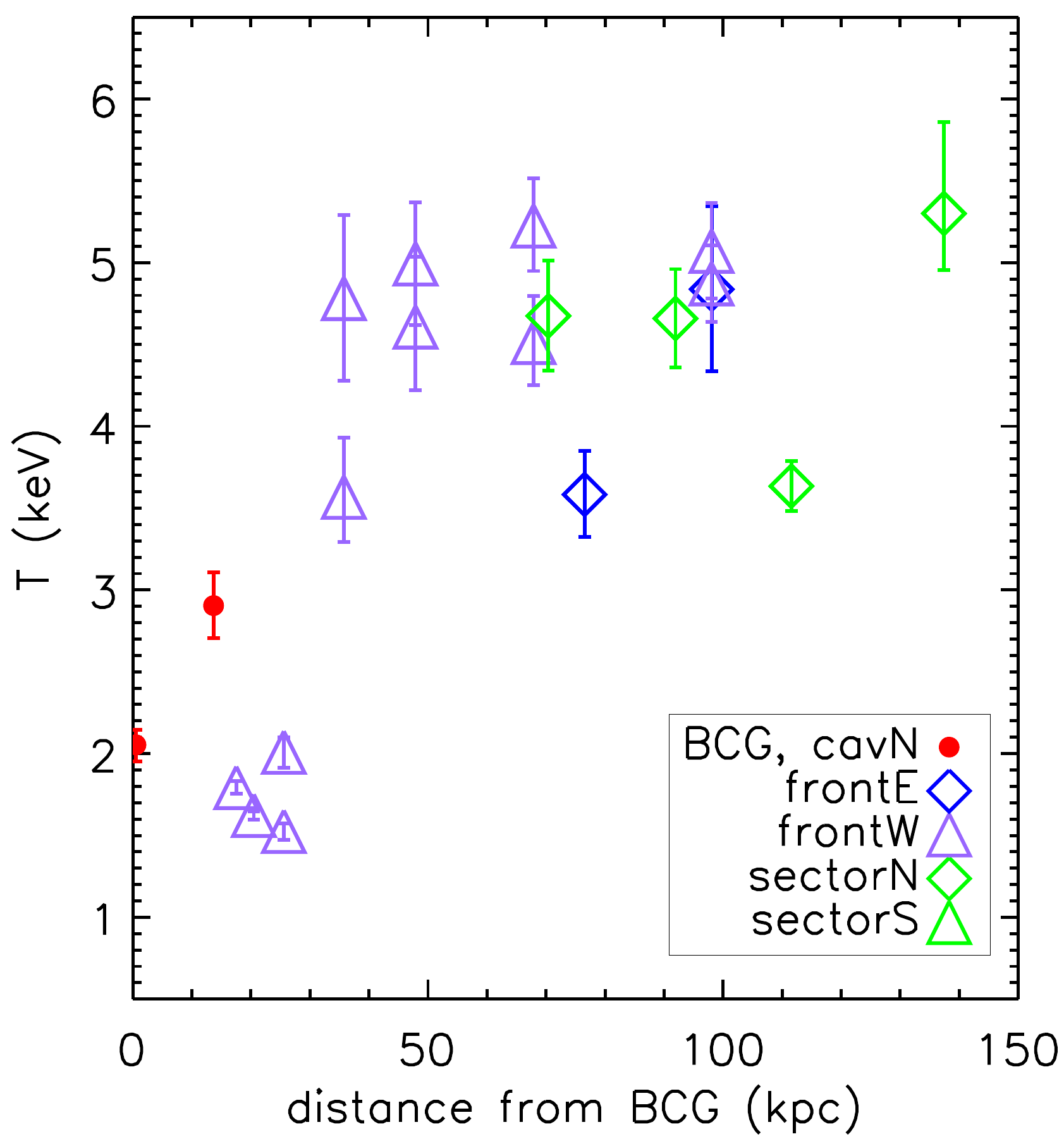,width=0.5\textwidth}
\caption{Best-fit temperatures, from the 1T model for the selected regions listed in Table~\ref{tab:spectra}, as a function of the distance of the region where the spectrum has been accumulated from the BCG. 
} \label{fig:1Tdist} \end{figure}

In Fig.~\ref{fig:1Tdist}, we plot the best-fit gas temperature, estimated with a single thermal (1T) model, as a function of the distance of the centre of these regions from the brightest cluster galaxy. 
The typical radial structure of a cooling core cluster with an ambient gas temperature of about 5 keV appears.

We discuss below the three most interesting regions: the front in surface brightness at west, another possible front at east, and an anisotropic elongated structure overabundant in metals compared to the neighbouring ICM.
The map of residuals and a {\it mild decrement} in the surface brightness profile also suggest the presence of an X-ray cavity at $10$ kpc north of the BCG and of an even weaker cavity to the south of the BCG.
Given the poor counts statistic available in this area, we can only estimate a gas temperature of about 3 keV with metallicity of 0.65 times the solar value (0.9 when a 2T thermal model is adopted; see Table~\ref{tab:spectra}).

In the remaining regions, a  single-phase gas provides a good description of the thermal emission in the northern and southern sectors (labelled {\it sectorN} and {\it sectorS} in Table~\ref{tab:spectra}), with the region at the south that presents a smoother temperature distribution around 4.7 keV.

\begin{table*}
\centering
\input{1T_2T_paper.tex}
\caption{Spectral analysis of the regions of interest by adopting one and two absorbed {\tt apec} models in the energy range $0.6-7$ keV. 
The F-test column provides the significance level (high values mean 
high significance) of the improvement in the fitting statistic for the three additional parameters introduced with a second thermal model.
The elemental abundances are given with respect to the solar values from Anders \& Grevesse (1989). 
The distribution of the 1T spectral estimates as a function of the distance from the BCG is plotted in Fig.~\ref{fig:1Tdist} . 
} \label{tab:spectra}
\end{table*}

\subsection{The cold front west of the BCG}

Located at about 23 kpc from the BCG, this front shows a clear decrement in the surface brightness (Fig.~\ref{fig:mapt}) and a rise in the temperature estimate from about 2 keV to 5 keV once a single thermal model is adopted. However, the gas inside the discontinuity (labelled {\it frontW-in} in Table~\ref{tab:spectra}; see circular region in Fig.~\ref{fig:mapt} that includes {\it frontW-inN} and  {\it frontW-inS})
shows a statistically high significance for the need of a second thermal component. 
This gas, therefore, seems multiphase, with a cool thermal component at 1.5 keV and a second component hotter than 2.6 keV (at $1 \sigma$). 

We performed detailed spatial and spectral analysis in the region around this front.
In proximity of the edge in the surface brightness profile, we modelled the gas density with a broken power law
\begin{equation}
n_{\rm e}(r)=\left\{
\begin{array}{ll}
  n_{\rm e, in}\left({{r}\over{R_{\rm f}}}\right)^{-\beta_{\rm in}}, &  r<R_{\rm f},\\
  n_{\rm e, out}\left({{r}\over{R_{\rm f}}}\right)^{-\beta_{\rm out}}, &  r>R_{\rm f},
\end{array}
\right.
\label{eqn:density}
\end{equation}
where $R_{\rm f}$ indicates the location of the front.
We projected this model along the line of sight and fit it to the surface brightness profile. 
We measured a ratio between the gas densities at $R_{\rm f}$ of $3.01 \pm 0.10$, when the sector between 335 and 30 degrees (Cartesian X-axis = $0^{\circ}$; north is $90^{\circ}$) is considered. Once we zoom into this region and define two additional sectors, one pointing to the north (0--80 degrees) and the other southward (295--0 degrees), we estimate a ratio of $3.07 \pm 0.06$ and $3.67 \pm 0.10$, respectively.
Analysing the corresponding spectra extracted from boxes aligned along the front, we measure
$T = 4.14 \pm 0.38$ keV and $4.58 \pm 0.42$ in the northern and southern sector, with metal abundance of $0.58 \pm 0.21$ and $0.90 \pm 0.29$ solar, respectively; when we fix these values as second components contributing to the total emission in the internal part and leave free the normalization, we obtain $T = 1.71 \pm 0.14$ keV and $1.20 \pm 0.05$ in the northern and southern sector, with metal abundance of $0.57 \pm 0.16$ and $0.32 \pm 0.10$ solar, respectively.
These measurements correspond to a ratio between the temperatures of $2.42 \pm 0.38$ and $3.82 \pm 0.37$ for the sectors to the north and south, respectively, which agrees well with the values observed in the density discontinuity.
Combining the estimates of the jumps in density and temperature, we can evaluate the jump in the pressure profile to be
$0.79 \pm 0.13$ in the north, $1.04 \pm 0.11$ in the south, consistent with unity.
Thus we confirm that the front is indeed a contact discontinuity and not a shock.

\begin{figure*}
\epsfig{figure=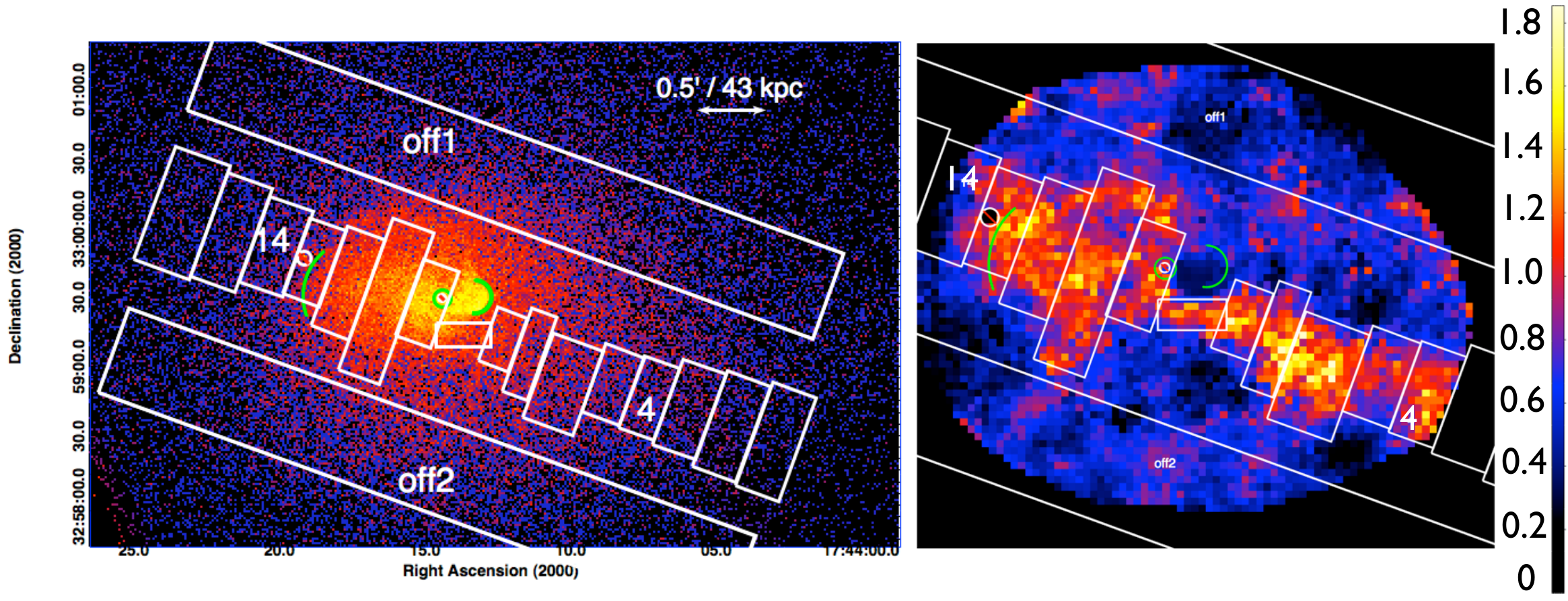,width=\textwidth}
\caption{Regions of interest for the spectral analysis plotted over {\it (left)} the exposure-corrected image and {\it (right)} the abundance map indicating the spectroscopically estimated values of metallicity following the colour bar on the right.  
The regions labelled from 1 to 16 (here, only 4 and 14 are indicated for the sake of clarity) have the measured metal abundance plotted in Fig.~\ref{fig:metprof}.
The BCG (green circle) and the two fronts (green arcs) are indicated.
} \label{fig:mapz} \end{figure*}

\subsection{A possible front at east and a speculative shock at south/west of the BCG}

At about 80 kpc from the BCG, we notice a decrement in the surface brightness profile extracted in the sector between 130 and 200 degrees and an increase in the gas temperature from 3.7 to 5 keV, with the metallicity that decreases by a factor of 2. This gas appears isothermal, statistically speaking.

The fit with a broken power law in the gas density distribution provides a measure of the discontinuity of $1.49 \pm 0.03$ (see Fig.~\ref{fig:fronts}).
Across the front we measure a (projected) temperature $2.16^{+0.46}_{-0.14}$ and $5.11 \pm 0.35$ keV inside and outside the front, respectively, with an estimate of the pressure jump $P_{\rm out} / P_{\rm in}$ of $1.59 \pm 0.36$.

Another front is probably detected at the region of higher temperature located in the temperature map shown in Fig.~\ref{fig:mapt} at about 0.7 arcmin ($\sim 60$ kpc) from the BCG along the direction to south-west (between 305 and 335 degrees in our reference frame).
Along this sector, we measure an increasing temperature up to $6.7 (\pm 1.9)$ keV in the annuli just before the break in the surface brightness profile shown in Fig.~\ref{fig:fronts}, with a lower temperature ($5.4 \pm 0.8$ keV) behind it. This jump in temperature ($\sim 1.2 \pm 0.4$) would imply a gas density jump of about 1.3, for a Mach number$\approx 1.2$ shock discontinuity following the Rankine-Hugoniot conditions (see, e.g., Markevitch \& Vikhlinin 2007).  The predicted ratio between the gas density before and after this shock matches the observed value of $1.4 \pm 0.5$ well. However, the large statistical uncertainties associated to these values weaken the significance of such detection. 

\begin{figure*}
\hbox{
 \epsfig{figure=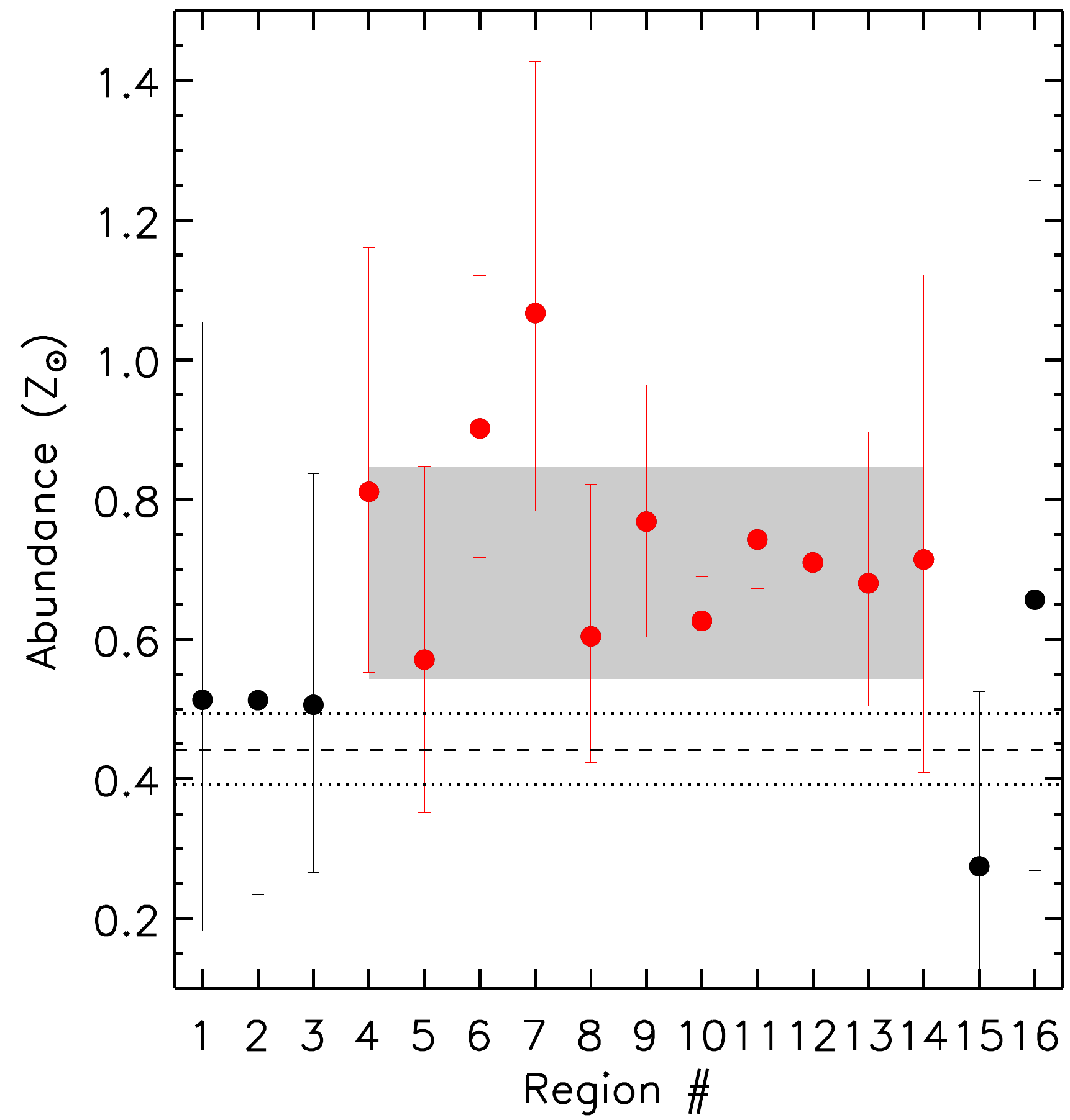,width=0.5\textwidth}
  \epsfig{figure=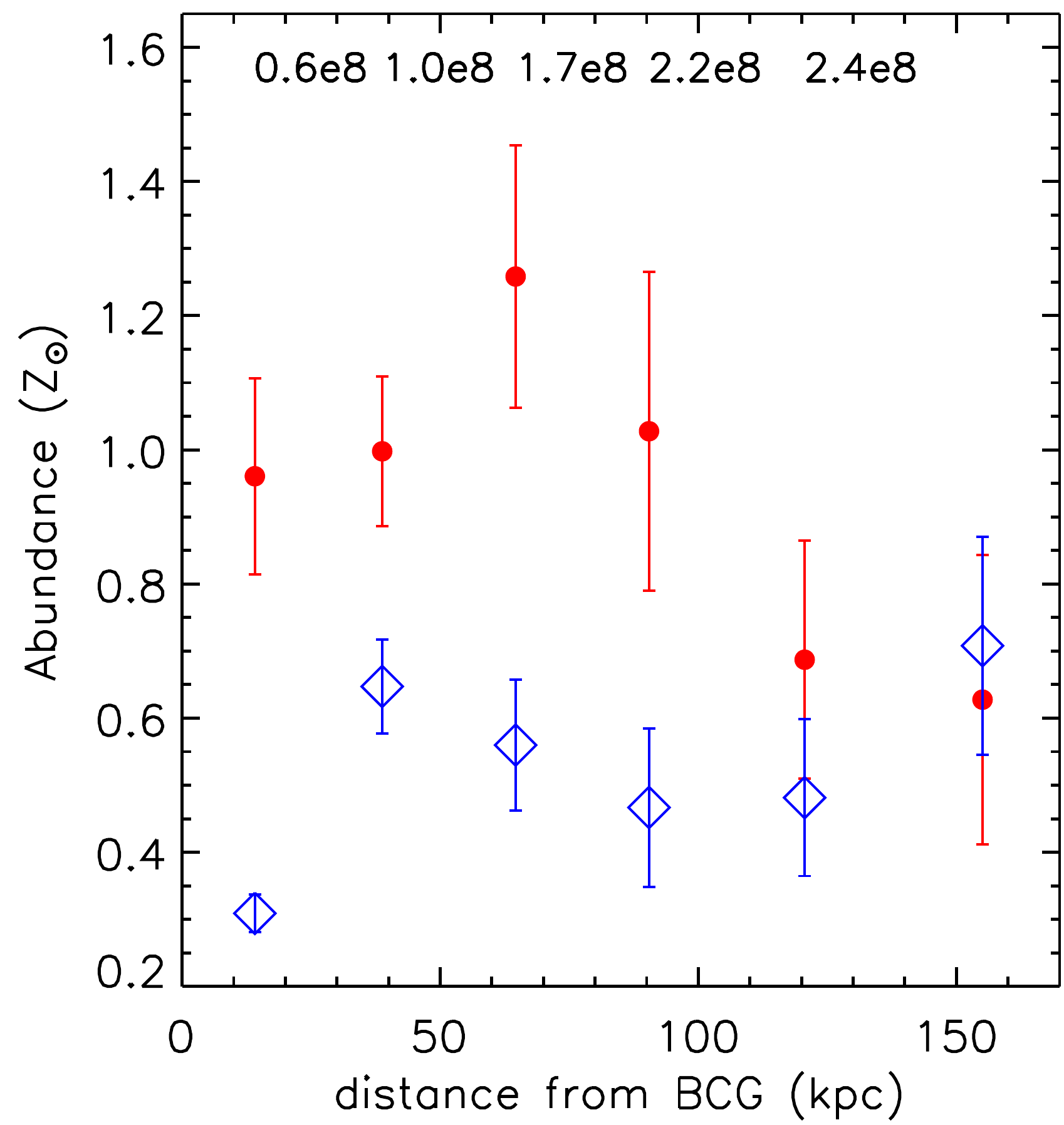,width=0.5\textwidth}
}
\caption{Distribution of the metal abundance of the ICM along the structure described in Fig.~\ref{fig:mapz}. ({\it Left}) Estimates along the regions labelled from 1 to 16.  
The apparently coherent structure corresponds to the regions between 4 and 14 (red points) where the measures of the metal abundance are systematically higher than the values measured in the surrounding gas by about 50\%.   The difference between the weighted averages of the abundance values along this structure and in the ``off'' region (a combination of  ``off1'' and ``off2'' regions) used as reference is $\sim1.6 \sigma$ when the scatter is considered (shaded region and dotted lines, respectively) and increases to $4.0 \sigma$ when the error on the mean is used.
The dimension of the elongated structure is about $(3.2 \times 1)$ arcmin$^2$ $\approx (276 \times 86)$ kpc$^2$.
({\it Right}) Spectral measurements as function of the distance from the BCG. 
The measurements were done along two different directions, one along the elongated structure (red dots) and the other almost orthogonally to it (blue diamonds). 
The two values match at about 150 kpc. At the top, the excess in iron mass in unit of solar mass as a function of increasing radius is indicated.
The elemental abundances are given with respect to the solar values from Anders \& Grevesse (1989).  
} \label{fig:metprof} \end{figure*}

\subsection{The metal-rich elongated structure}

The abundance map in Fig.~\ref{fig:mapz} highlights an elongated structure with metal abundance higher by about 50\% than the mean value estimated in the surrounding regions. 
A detailed spectral analysis shows that a single-phase gas is a good model for the spectra extracted from boxes with dimensions of about $30 \times 20$ arcsec$^2$ and indicates a clear enhancement in the metal abundance in a structure with size $\approx (276 \times 86)$ kpc$^2$ and an orientation at an angle of 153 degrees from the Cartesian X-axis (see Fig.~\ref{fig:mapz} and \ref{fig:metprof}). This metal-rich elongated structure has an orientation very similar to both the overall distribution of the X-ray light ($\sim 171$ degrees) and the direction connecting the west and east fronts ($\sim 167$ degrees; all these angles are measured anti-clockwise from the Cartesian X-axis). This structure also has an associated enhancement of the X-ray surface brightness as evaluated from the map of the residuals.

\begin{figure*}
\centering
 \epsfig{figure=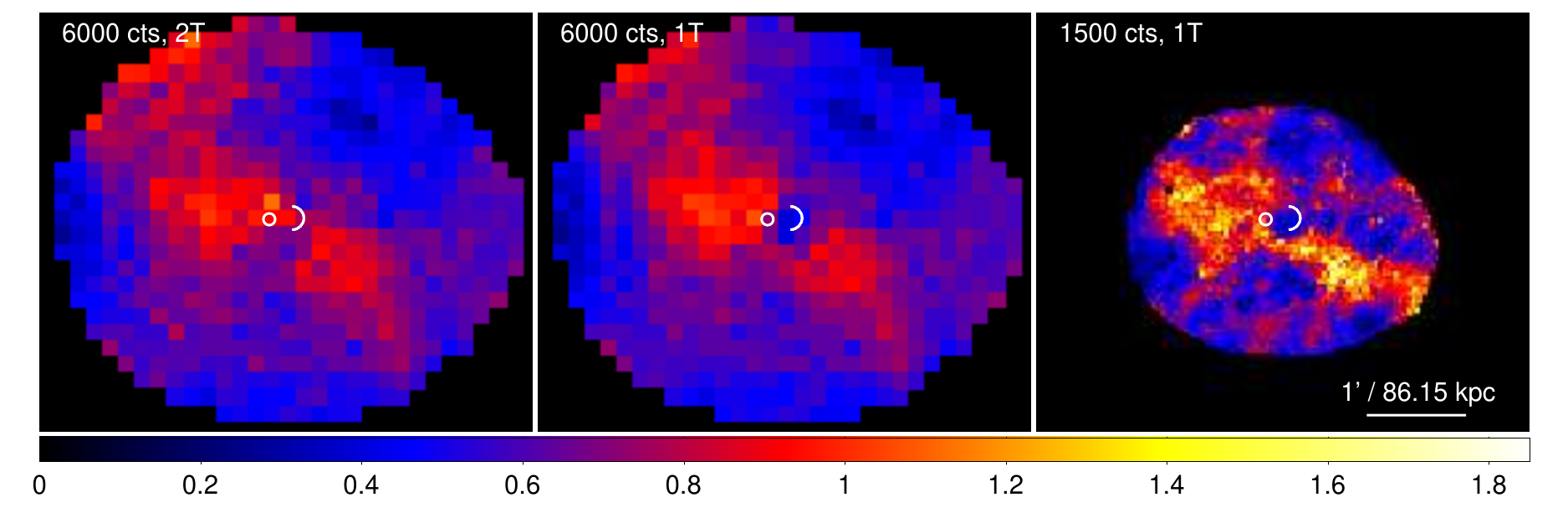,width=\textwidth}
\caption{Maps of the measurements of the metal abundance as obtained with ({\it left}) a 2T model and requiring 6000 net count, ({\it centre}) a single temperature model and the same number of net counts, ({\it right}) the 1-T model and 1500 net counts (i.e. the same abundance map shown in Fig.~\ref{fig:specmap}). 
The BCG (white circle) and the front to the west (white arc) are indicated.
} \label{fig:1T2T} \end{figure*}

An additional abundance map has been obtained considering 6000 (instead of 1500) net counts to allow the modelling with a 2T thermal model with the abundance tied between the two components.
This analysis allows us to appreciate the impact of a 2T model on the estimate of the ICM metallicity in the cluster and to highlight the iron-bias due to a single thermal modelling of a multi-phase gas.
In Fig.~\ref{fig:1T2T} we show the comparison between the 2T abundance map obtained with 6000 count, the same map fitted
with a single temperature model, and the original high-resolution 1T abundance map. The only remarkable effect of the 2T fits is to remove the biased low abundances in the cool core, whereas the metallicity values are unaffected elsewhere.

In the right panel of Fig.~\ref{fig:metprof} we show the metal abundance profiles extracted along two different directions, one following the elongated structure and the other almost orthogonal to it.
We quantify (and quote in the plot) the difference in metallicity as an excess in iron mass associated to the metal-rich structure.
This excess is measured as $\Delta M_{\rm Fe} = A_{\rm Fe} \, m_H \, Z_{\odot} \, (4 \pi \times 10^{14})^{0.5} D_{\rm lum} /(1+z) \sum_i \Delta A_i K_i^{0.5} V_i^{0.5}$, where  
$A_{\rm Fe}$  is the atomic weight of iron; $m_H$ is the proton mass; the luminosity distance $D_{\rm lum}$ of 342.7 Mpc at the cluster redshift is adopted. The sum is done using the spectroscopic best-fit values of abundance $A_i$ and normalization $K_i$ (from the adopted {\tt apec} model) and assuming a cylindrical volume $V_i$ (with height given from the extension of the region considered and a radius of the base of 43 kpc; no significant change is obtained by assuming slightly different geometries). 
The estimated value of $\Delta M_{\rm Fe} \approx 2.4 \times 10^8 M_{\odot}$ within 150 kpc is fully consistent with the iron production expected from the BCG in few Gyrs (e.g. De Grandi et al. 2004, B\"oringher et al. 2004).
For example, assuming a BCG B-band luminosity of $1.3\times 10^{11}$ L$_{B,\odot}$, we find that in the last 5 Gyr 
the SN-Ia+stellar winds inject $\sim 1.7\times 10^8$ M$_\odot$ of iron in the ICM. For this estimate we adopt a 
current SNIa rate of 0.15 SNu, with a time dependency $SNu(t) \propto t^{-1.1}$ and a mean stellar
metallicity of 1.6 solar (see B\"oringher et al. 2004 for further details). The SN-II are not expected to be important
contributors of iron in the last few Gyrs, given the estimated star formation rate ($\sim 3.7$ M$_\odot$ yr$^{-1}$,
O'Dea et al. 2008).

\section{Discussion and conclusions}

The system \obj\ is a massive galaxy cluster (ambient temperature of 4.7-5 keV) where the X-ray emitting gas is characterized by an extended region at low entropy.
This ICM is probably sloshing in the gravitational potential, with peculiar imprints such as two well-defined cold fronts (one at 23 kpc west of the BCG, the other at 80 kpc east of the BCG), a speculative Mach number$\approx 1.2$ weak shock (at about 60 kpc south/west of the BCG) and a metal-rich elongated structure with size 280 kpc times 90 kpc, almost aligned both to the total X-ray light distribution and to the relative location of the fronts. This elongated structure has a measured metallicity higher than the mean values in the neighbouring regions by about 50 \%, with a significance of about 
$\sim 2 \sigma$ when the scatter around the mean is considered and a corresponding  excess in iron mass of about $2.4 \times 10^8 M_{\odot}$ (see Fig.~\ref{fig:mapz} and \ref{fig:metprof}).
A plausible cavity a few tens of kpc north of the BCG, and a weaker counter-part to the south, can be either the product of the activity of the central AGN or the apparent product of the enhanced emission in the neighbouring regions caused by the ongoing sloshing activity. If real, these cavities are probably connected to breaks in the jets/outflows and are more like to be destroyed from the central turbulence/sloshing, suggesting that they are not the main source of the feedback acting in \obj.

Few plausible scenarios have been suggested to explain the overall picture of this system.
In one of these scenarios, the cool core with active star formation undergoes a minor-merger, along a projected direction with an angle of about 170 degrees (measured anti-clockwise from a Cartesian X-axis). This should induce gas sloshing in the underlying cluster's potential. 
Roediger et al. (2011; see also Ascasibar \& Markevitch 2006), for instance, use hydrodynamical simulations to show that the evolution of the metal distribution is tightly coupled to the overall dynamics. They obtain metallicity structures similar to the temperature structures. However, they also prove that the sloshing alone cannot, for instance, redistribute a metal distribution with a steep peak originating from stellar mass loss of the central galaxy into a flatter one (as for the case of M87). To produce that, they require additional processes such as diffusion induced from turbulence (e.g. Rebusco et al. 2005, 2006) caused by AGN-inflated buoyant bubbles or repeated minor-mergers, for example. Active Galactic Nuclei activity, in particular, can act in a complementary way to the gas sloshing, the former being more effective in the very cluster centre, the latter at larger radii. They conclude that the observed metal distribution, even across the cold fronts, depends strongly on the initial (unknown) distribution; sloshing affects it towards the cold fronts and broadens it on larger scales.
It is, therefore, difficult to outline a picture of the formation of the metal elongated structure, whether it was already in situ as a product of some major-merger that formed \obj\ or if it was generated during the last minor-merger that initiated the gas sloshing. 
In the second scenario, the cool core of the small merging group, after a first passage during which metals are deposited along the elongated structure and the front is formed to east, now returns to the centre of the potential well, around the position of the BCG, generating the front to west. It also continues to pollute the surrounding ICM with metals.
In this case, it would be the first example, to our knowledge,  in which a moving cool core can be traced not only with the position of the cold fronts but also with the enhanced metal distribution, with respect to the ambient gas, along the direction of the sloshing activity.

An alternative scenario implies a mechanically dominated outflow from the central AGN that uplifts the metals produced in the BCG up to 100s kpc.
Kirkpatrick et al. (2011) find that the radial extent of a metal outflow scales with the jet power as $R_{Fe} = 58 \left( P_{jet} / 10^{44} {\rm erg\, s}^{-1} \right)^{0.42}$ kpc, with a rms scatter around the fit of about 0.5 dex.
From Fig.~\ref{fig:mapz}, we estimate the maximum radius $R_{Fe}$ of $\approx 150$ kpc at which a significant enhancement in metallicity has been detected. It corresponds to a jet power of about $10^{45}$ erg s$^{-1}$, which is consistent with the results of simulations showing that metals originating mainly from SnIa explosions occurring in the cD galaxy are easily transported along the jet-axis up to 150-200 kpc when the AGN is very active (see e.g. Gaspari et al. 2011a, b).  
This activity would naturally explain the evidence that: (i) the metal distribution is highly anisotropic, with a bipolar symmetric pattern; 
(ii) at larger radii, the metal distribution expands laterally, forming two or more lobes of conical geometries, since ambient pressure diminishes (and thus also collimation); (iii) eastern and western arms show similar metal contrast, slightly more metallic along the spine (observation that is indeed difficult to reproduce by sloshing); (iv) cold/warm dense gas uplift is very common for mechanical feedback, especially within tens kpc, and this could provide a simple explanation for the clear cold front near the centre (along with the small X-ray peak shift). 
Overall, we argue that the previous AGN outflow event has uplifted cold dense gas in the inner few tens of kpc (more shifted towards west), creating the cold front. At the same time, it has enriched the hotter gas up to 300 kpc. The direction of the bipolar outflow has probably been distorted in an S-shaped geometry near the kpc base, because of denser clumps (possibly creating fragmented bubbles like the ones marginally detected to the north and south of the BCG).
For the outflow scenario to work in the presence of the elongated metal-enrich structure, the relatively central ICM should have acquired $\approx 2\times 10^8 M_\odot$ of iron before the occurrence of the AGN outburst. The iron reservoir may pile up in the core during a more quiescent Gyr evolution, or be the result of the complex dynamics associated with previous feedback events.
Turbulence and bulk motion may alter the metallicity structure, inducing a more isotropic distribution in iron abundance, although this is balanced by the recurrent outflows (with typical duty cycle $\approx 100$ Myr). 
The detected weak shock could be, therefore, a by-product of a recent period of activity (bubbles are not essential to produce AGN feedback). 
Nevertheless, it is difficult to trace the activity of previous events in X-ray surface brightness maps, especially at $r > 100$ kpc 
(the underlying profile steeply decreases with radius and bubbles/shocks quickly decay). 
For example, Figs. 11 (panel h) and 14 (panel b) in Gaspari et al. (2011a) show that the X-ray surface brightness map is very regular up to 200 kpc, while the metal anisotropy is still present. 

A follow-up in radio wavebands (3h EVLA observation at 1.4 GHz, PI: Gitti, are in progress) and with $H\alpha$ resolved imaging would help to characterize the cool core emission and its interplay with central AGN activity, and would also better define the occurrence of thermal instabilities, with layers of H$\alpha$ emitting gas that form from warm gas condensing out of the hot phase and that surround the dense core of molecular gas (e.g. Gaspari et al. 2012 and references therein).

\section*{Acknowledgments}
We thank the anonymous referee for the useful comments that improved the presentation of the work.
We acknowledge the financial contribution from contract ASI-INAF I/023/05/0 and I/088/06/0.
SE acknowledges financial support from \chandra\ grant GO0-11136X and from FP7-PEOPLE-IRSES-CAFEGroups 
(grant agreement 247653).
FG acknowledges financial contributions from the Italian Space Agency through 
ASI/INAF agreement I/032/10/0 for XMM-Newton operations.

\end{document}

%% file: 1T_2T_paper.tex
\begin{tabular}{cccccccc} \hline
Region & $1T$ (keV) & $A$ ($Z_{\odot}$) & C-stat/DOF & $2T$ (keV) & $A$ ($Z_{\odot}$) & C-stat/DOF & F-test \\ \hline
BCG-noAGN & $2.05^{+0.09}_{-0.10}$ & $0.37^{+0.12}_{-0.09}$ & 186.4/193 & $2.46^{+0.21}_{-0.21}$, $0.73^{+0.11}_{-0.15}$ & $1.01^{+0.49}_{-0.32}$ & 178.2/191 & 0.986 \\
cavN & $2.90^{+0.20}_{-0.20}$ & $0.65^{+0.18}_{-0.15}$ & 191.2/238 &  \\
frontE-in & $3.58^{+0.16}_{-0.16}$ & $1.04^{+0.24}_{-0.21}$ & 233.5/263 & $1.37^{+0.77}_{-0.34}$, $4.11^{+1.62}_{-0.48}$ & $1.04^{+0.35}_{-0.25}$ & 230.7/261 & 0.793 \\
frontE-out & $4.84^{+0.50}_{-0.50}$ & $0.53^{+0.13}_{-0.23}$ & 273.8/276 & \\
frontW-in & $1.62^{+0.02}_{-0.03}$ & $0.26^{+0.03}_{-0.03}$ & 377.9/294 & $3.82^{+3.31}_{-0.83}$, $1.25^{+0.05}_{-0.07}$ & $0.43^{+0.14}_{-0.11}$ & 295.1/292 & 1.000 \\
frontW-inN & $2.01^{+0.09}_{-0.10}$ & $0.40^{+0.12}_{-0.09}$ & 127.7/188 & $4.31^{>10}_{-1.70}$, $1.53^{+0.14}_{-0.25}$ & $0.47^{+0.25}_{-0.14}$ & 122.3/186 & 0.981 \\
frontW-outN & $3.57^{+0.36}_{-0.27}$ & $0.63^{+0.23}_{-0.18}$ & 221.0/226 &  \\
frontW-out2N & $4.61^{+0.43}_{-0.39}$ & $0.46^{+0.19}_{-0.15}$ & 271.5/279 & $5.03^{>10}_{-2.51}$, $1.41^{>10}_{-0.73}$ & $0.56^{+0.26}_{-0.20}$ & 270.6/277 & 0.363 \\
frontW-out3N & $4.51^{+0.28}_{-0.26}$ & $0.44^{+0.12}_{-0.10}$ & 361.6/356 &  \\
frontW-out4N & $5.07^{+0.30}_{-0.29}$ & $0.39^{+0.13}_{-0.12}$ & 330.5/367 &  \\
frontW-inS & $1.53^{+0.05}_{-0.05}$ & $0.24^{+0.05}_{-0.05}$ & 246.3/188 &  \\
frontW-outS & $4.78^{+0.51}_{-0.50}$ & $0.81^{+0.35}_{-0.27}$ & 204.4/228 &  \\
frontW-out2S & $4.99^{+0.37}_{-0.38}$ & $0.67^{+0.22}_{-0.18}$ & 260.9/300 & $6.34^{+4.06}_{-1.18}$, $1.73^{+1.60}_{-0.38}$ & $0.78^{+0.30}_{-0.24}$ & 258.9/298 & 0.676 \\
frontW-out3S & $5.22^{+0.29}_{-0.28}$ & $0.55^{+0.13}_{-0.12}$ & 388.0/358 &  \\
frontW-out4S & $4.87^{+0.23}_{-0.23}$ & $0.56^{+0.11}_{-0.10}$ & 384.9/390 &  \\
sectorN-in & $3.63^{+0.15}_{-0.15}$ & $0.59^{+0.09}_{-0.08}$ & 331.3/364 & $7.49^{>10}_{-5.06}$, $2.91^{+0.45}_{-0.57}$ & $0.73^{+0.13}_{-0.12}$ & 329.4/362 & 0.650 \\
sectorN-out & $5.30^{+0.56}_{-0.34}$ & $0.42^{+0.16}_{-0.14}$ & 284.0/351 &  \\
sectorS-in & $4.67^{+0.34}_{-0.33}$ & $0.46^{+0.13}_{-0.11}$ & 326.0/336 &  \\
sectorS-out & $4.66^{+0.30}_{-0.30}$ & $0.33^{+0.11}_{-0.10}$ & 341.1/359 & \\
\hline \end{tabular}